\documentclass[]{article}

\pdfoutput=1

\usepackage[square]{natbib}
\usepackage{times}
\usepackage{amsmath}
\usepackage[english]{babel}

\usepackage{color}
\usepackage{tikz}
\usepackage{hyperref}
\usetikzlibrary{arrows,calc}
\hypersetup{colorlinks, breaklinks,
  citecolor=[rgb]{0.0,0.3,0.3},
  linkcolor=[rgb]{0.5,0.5,0.5}}
          
\usepackage{authblk}

\renewcommand{\cite}[1]{\citet{#1}}

\newcommand{\cmnt}[1]{}
\newcommand{\ignorewhencounting}[1]{#1}

\newcommand{\vex}[1]{\boldsymbol{#1}}
\usepackage[babel]{csquotes}
\usepackage{booktabs}

\newcommand{\setmeter}[2]{\ensuremath{%
\vcenter{\offinterlineskip
\halign{\hfil##\hfil\cr
$\scriptstyle#1$\cr
\noalign{\vskip1pt}
$\scriptstyle#2$\cr}
}}%
}

\title{Towards computer-assisted understanding of dynamics in symphonic music\footnote{To appear in IEEE Multimedia (\url{https://www.computer.org/multimedia-magazine/}). \copyright\ IEEE}}

\newcommand{\ts}[1]{\textsuperscript{#1}}
\author{Maarten Grachten\ts{1}, Carlos Eduardo Cancino-Chac\'on\ts{2}, Thassilo Gadermaier\ts{2} and Gerhard Widmer\ts{1,2}}
\affil{\ts{1} Department of Computational Perception, Johannes Kepler University, Linz, Austria}
\affil{\ts{2} Austrian Research Institute for Artificial Intelligence, Vienna, Austria}
\date{}

\begin{document}

\maketitle

\begin{abstract}
Many people enjoy classical symphonic music. Its diverse instrumentation makes for a rich listening experience.
This diversity adds to the conductor's expressive freedom to shape the sound according to their imagination. As a result, the same piece may sound quite differently from one conductor to another.
Differences in interpretation may be noticeable subjectively to listeners, but they are sometimes hard to pinpoint, presumably because of the acoustic complexity of the sound.
We describe a computational model that interprets dynamics---expressive loudness variations in performances---in terms of the musical score, highlighting differences between performances of the same piece.
We demonstrate experimentally that the model has predictive power, and give examples of conductor ideosyncrasies found by using the model as an explanatory tool.
Although the present model is still in active development, it may pave the road for a consumer-oriented companion to interactive classical music understanding.

\end{abstract}


\section{Introduction}\label{sec:introduction}
When you ask visitors of a classical concert what it is that makes attending a concert worthwhile you will get a variety of answers, but some likely reasons are that a good performance affects you emotionally,
and can be so immersive that you forget the world around you~\citep{roose08:_many_voiced_unison}.
The related question what it is that makes a good performance, is equally ambiguous, and ultimately depends on personal taste.
Nevertheless, it has been long known that music is more engaging when it is played \emph{expressively}.
A straight-forward mechanical reproduction of a written musical score, as a computer would produce it, typically sounds dull, and to the trained listener it may sound odd, or even plain wrong.

The performance of a piece of music can be called expressive when it conveys information to the listener that a literal, mechanical rendition of the score would not.
The information may be an affective quality (for instance, the listener may perceive a piece as being performed \emph{solemnly}, or \emph{joyfully}), but the performance may also express structural information about the music (for instance, the listener may notice from the way the music is performed that a musical phrase is coming to an end).

Musicians convey such information by varying the way they perform the written score.
Among \emph{tempo}, and \emph{articulation}, one of the more salient expressive aspects of the performance is \emph{dynamics}--variations in loudness of the performance for the purpose of musical expression.
By varying these parameters during the interpretation of the piece, musicians make a performance sound more natural, and alive.
Careful control of these parameters also allows musicians to create \emph{phrasing} in the music, producing a perception of coherence in the music over longer time spans.

Appreciation of the music by listeners is facilitated by their familiarity with the piece, and music understanding in general.
This is reflected in a desire for information, expressed by concert-goers, about the music they are to hear in the concert~\citep{melenhorst15:_put_concer_atten_spotl}.
Although it is relatively straight-forward to obtain biographical or historical information about a composer or a piece, through web or library search, there are few facilities that help listeners become familiar with a the musical details of a piece, or a specific performance.
Apart from synchronized musical scores available nowadays in music videos from online services such as \emph{Youtube}, a notable step in that direction is the iPad Magazine by the Dutch Royal Concertgebouw Orchestra, RCO Editions, in which recorded performances can be played back in sync with the musical score.
The data used in the experimental evaluation of the model presented here originates (mostly) from a collaboration between RCO and the Austrian Research Institute for Artificial Intelligence (OFAI) to produce the score-synchronization in the RCO Editions.

Although musical expression plays an important role in the musical experience, to date there are virtually no facilities for the interested listener to learn more about the expressive aspects of music.
A tool that can attribute variations in the expressive quality of a performance to factors like performance directives (like \emph{crescendo}, \emph{diminuendo}, and \emph{fermata}), and other aspects of the written score, may elucidate expressive intentions of the conductor to listeners, thereby stimulating their engagement with, and understanding of the music.
In this way, it addresses the needs of (actual or potential) classical music listeners, who in a user study ``...expressed interest in the structure of the music, the composer's intention, the conductor's interpretation, and the discovery of style differences in comparison to recordings.'' \citep[Sec 7.2]{melenhorst15:_put_concer_atten_spotl}. 
Such a tool may be part of an \emph{active music listening interface}~\citep{goto07:_activ_music_listen_inter_based_signal_proces} for classical music such as the integrated prototype\footnote{\url{http://beta.phenicx.com/}} of the PHENICX project~\citep{liem2015phenicx}.

An important question to be addressed in the development of an end user tool for understanding dynamics is what level of information is appropriate for the end user.
Unexperienced listeners may benefit most from a simple approach, such as merely highlighting the parts of a piece where two performances differ substantially. For such use cases, where the need for an explanatory model is reduced, a more descriptive approach like that of \cite{liem15:_compar_analy_orches_perfor_recor} may be useful.
Musically trained listeners however, may be interested in further details, and may be helped by a tool that during the playback of a performance, highlights aspects of the score that explain expressive peculiarities of the current performance with respect to a typical performance of the same piece.
Finally, a tool for musicologists would not only provide a qualitative explanation of differences in expressive interpretation between performances, but also allow for the use of these explanations in a comparative analysis of sets of performances in terms of expression, uncovering consistent expressive strategies of conductors, or grouping performances in terms of their expressive characteristics.

The purpose of the current article is to pave the way for such a facility.
More specifically, we present a computational model of dynamics in music, and show how such a model may help to understand the factors that contribute to dynamics.
Although the user requirements may vary considerably for the different use cases described above, we believe it is desirable to have a unified technological basis for computer-assisted understanding of dynamics at different levels of user expertise.
In this paper we focus on the capability of the proposed model to extract relatively detailed information from a performance, that we believe is most useful for expert users, like musicologists.
Although beyond the scope of this paper, our belief is that when the model can extract useful information for this class of users, appropriate summarization and selection of the extracted information can help the model cater to use cases involving less experienced users.

In the following, we give a brief overview of related work in computational modeling of musical expression, and continue to give a description of the model.
We report an experimental evaluation of the model, showing that it explains a considerable portion of the dynamics in recorded performances, on the basis of only the written score.
After this validation, we illustrate how the model may be used as a tool to visualize and explain differences in dynamics between performances.

\section{Computational modeling of musical expression}
Research into musical expression is ongoing, and our knowledge of underlying principles and mechanisms is far from complete.
In this light, computational modeling, machine learning, and data analysis methods have proven helpful in the study of musical expression, which largely relies on tacit knowledge by the musician.
For example, \cite{canazza03a} derive a mapping between sensorial expressive adjectives 
 and acoustic attributes of the performance.
\cite{widmer03} uses machine learning to infer simple rules from a large set of performances of Mozart piano sonatas, linking (for instance) expressive timing patterns to rhythmic characteristics of the music.
Another approach is taken by \cite{friberg06:_overv_kth_rule_system_music_perfor}, where a set of rules predict timing, dynamics and articulation, based on local musical context.
Using the rules from the system as macro-rules to model larger time scale performance trends,~\cite{bresin98} combined this symbolic model with a neural network to complement the macro-rules with micro-rules learned from a set of recorded performances.

Another computational model is proposed by \cite{grachten12:_linear}.
This model represents score information in terms of \emph{basis-functions} (BFs), and models dynamics as a linear combination of those basis-functions.
A distinctive feature of the model is that in addition to pitch and time information, it incorporates \emph{dynamic markings} such as \emph{crescendo}, and \emph{diminuendo} signs, by which the composer suggests a particular way of performing the piece.
Subsequent versions of the model have proven more effective by dropping the linearity constraint, allowing for non-linear combinations of basis-functions~\citep{chacon15}, and temporal dependencies~
\citep{grachten16:brnbm}.

\citet{liem15:_compar_analy_orches_perfor_recor} propose a method for comparing performances by a principal component analysis of audio spectrograms.
This method has the advantage that especially aspects like timbre can be studied, but it is purely descriptive of differences in the audio, rather than linking differences to particular information in the score, as in the method proposed here.

\section{A computational expression model for ensemble performance}
The tool we present here is based on the basis-function modeling (BM) approach mentioned above.
In this approach, the dynamics of a recorded ensemble performance over time is modeled as a combination of a set of basis-functions that describe the musical score.

\begin{figure}[t]
\begin{center}
\ignorewhencounting{
  \includegraphics[width=0.6\linewidth]{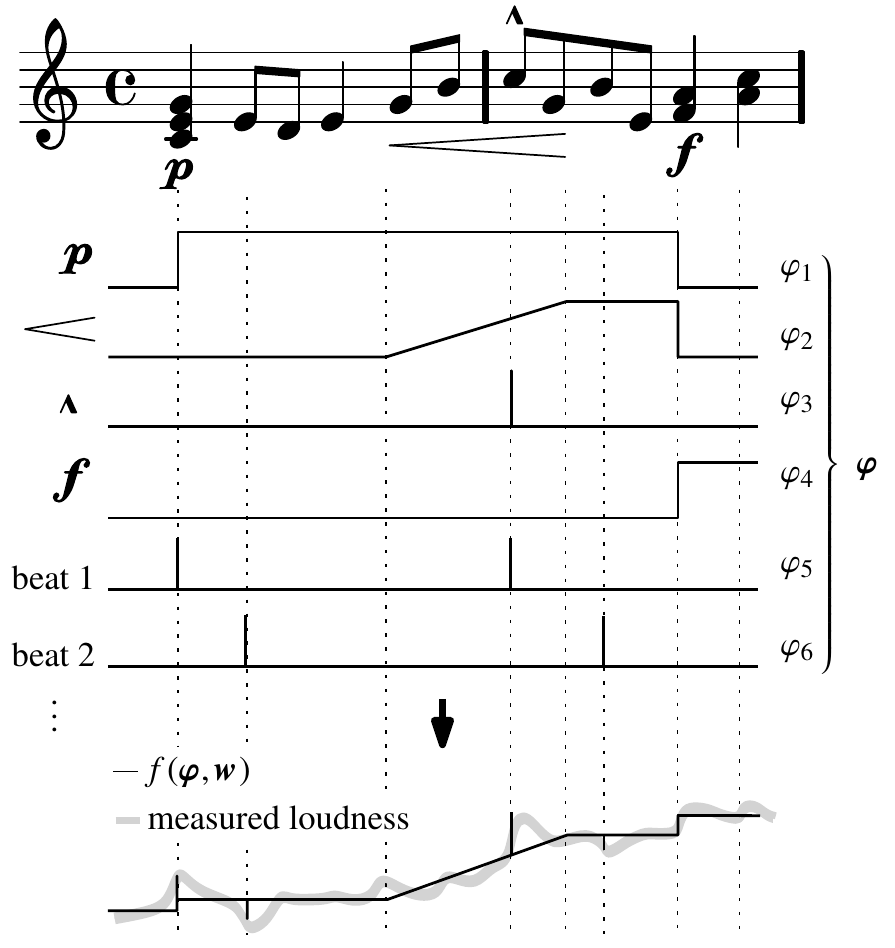}

\caption{Schematic view of dynamics as a function  $f(\vex{\varphi}; \vex{w})$ of basis-functions $\vex{\varphi} = (\varphi_1,\cdots,\varphi_K)$, representing dynamic annotations and metrical basis-functions
}
}
\label{fig:basisfunctiondiagram}
\end{center}
\end{figure}

Figure~\ref{fig:basisfunctiondiagram} illustrates the idea of modeling dynamics using basis-functions schematically.
Although basis-functions can be used to represent arbitrary properties of the musical score, the BM framework was proposed with the specific aim of modeling the effect of dynamic markings---hints in the musical score, to play a passage in a particular way.
For example, a \emph{\textbf{p}} (for \emph{piano}) tells the performer to play a particular passage softly, whereas a passage marked \emph{\textbf{f}} (for \emph{forte}) should be performed loudly.
Thus, \emph{\textbf{p}} and \emph{\textbf{f}} specify constant dynamic levels, and are modeled using a step-like function.
Gradual increase/decrease of dynamics (\emph{crescendo}/\emph{diminuendo}), indicated by right/left-oriented wedges, respectively, are encoded by ramp-like functions.
A third class of dynamic markings, such as \emph{marcato} (the \enquote{hat} over a note), or markings like \emph{sforzato} (\emph{\textbf{sfz}}), or \emph{forte piano} (\emph{\textbf{fp}}), indicate the accentuation of that note/chord.
This class of markings is represented through (translated) unit impulse functions.
A set of basis-functions $\vex{\varphi}$ is then combined in a function $f$, parametrized by a vector of weights $\vex{w}$, to approximate the dynamics measured from a recorded performance of the score.
The function $f$ may be a simple linear function, or a complex non-linear function, such as a neural network.
In either case, the vector of weights $\vex{w}$ determine how the basis-functions $\vex{\varphi}$ influence the estimated dynamics $f$, and it is through adjusting $\vex{w}$ that the model can be trained to predict dynamics, given a set of recordings.


The example basis-functions given above are schematic functions, representing a \emph{region} (step functions), a \emph{transition} (ramp functions), or the occurrence of some \emph{instantaneous event} (impulse functions).
Such functions are not limited to representing dynamic markings.
For example, figure~\ref{fig:basisfunctiondiagram} shows two impulse functions, representing the first and second beats in each bar, respectively.
Similarly (but not shown), transition functions effectively encode \emph{slurs}, and \emph{phrase marks}.
In addition to such schematic functions, basis-functions may encode numeric attributes of notes, such as their pitch, duration, and the number of notes sounding simultaneously.
By representing local, note level attributes, as well as mid (crescendi/diminuendi, slurs) and long (\emph{piano}/\emph{forte} marks, phrase-marks, repeats) range structures, the basis-function modeling approach allows for a uniform encoding of score information at different time scales. 


The basis-function approach is similar to the \emph{multiple viewpoint} (MV) representation of musical information proposed by~\citet{conklin1995multiple}, in the sense that both aim to provide a uniform way of representing diverse aspects of musical information for the purpose of modeling.
The main difference is that the latter has been conceived for use in Markov models that deal with discrete event spaces, and thus yields representations in terms of discrete values, whereas the former was designed for capturing quantitative relationships between score information and expressive parameters.

Before we address the question how the basis-functions can be used to model musical expression, we discuss how we represent dynamics as an expressive parameter, and how the ensemble scenario is different from the solo instrument scenario, since both issues have implications for the basis-function modeling approach.

\subsection{Measuring the dynamics of a music performance}
\label{sec:meas-dynam-music}
In earlier work~\citep{chacon15}, we have restricted the model to solo piano recordings, available in the form of precise measurements of the piano key movements, using a B\"osendorfer computer-monitored grand piano.
In such recordings, the recorded \emph{hammer velocity} of the piano keys is a direct representation of the dynamics, revealing how loud each individual note is played.
For acoustic instruments other than the piano, such precise recording techniques are not available, and therefore the dynamics of a complete symphonic orchestra cannot be measured in a similar way.
Another approach would be to record each instrument of the orchestra separately, and measure the loudness variations in each of the instruments.
This approach is not feasible either, because apart from the financial and practical barriers, the live setting in which orchestras normally play prevents a clean separation of the recording channels by instrument.

This means we are left with only a rudimentary measure of dynamics, namely the overall variation of loudness over time, measured from the orchestral recording.
Note that the loudness of a recording is affected by more than just dynamics.
Room acoustics, microphone positioning, and various processing steps during production, possibly including audio compression, and level-adjustments between instrument groups, may all affect the final loudness of the recording.
To some extent, such effects may be countered by normalizing the measured loudness per piece in terms of mean and variance.

We compute loudness of the recordings using the \cite{ebur128} loudness measure, which takes into account human perception (i.e., the fact that signals of equal power but different frequency content are not perceived as being equally loud) and is now the recommended way of comparing loudness levels of audio content in the broadcasting industry. To obtain instantaneous loudness values, we compute the \emph{momentary} variant of the measure on consecutive blocks of audio, using a block size and hop size of 1024 samples, using a 44100Hz samplerate.
Because we want to focus on \emph{variations} in loudness, rather than the overall loudness level and range, we subtract the mean and divide by the standard-deviation of the loudness values per recording.

Finally, in order to model loudness variations as a function of the score information, the performance of the piece must be aligned to the score. To that end we produce a synthetic audio rendering of the score, and align it to the recorded performance using the method described by~\cite{grachten13:_autom}.
Through the resulting score--performance alignment, the loudness curve computed from the recording can be indexed by musical time (such that we know the instantaneous loudness of the recording at, say, the second beat of measure 64 in the piece).
Note that the correctness of the alignment is a prerequisite for the explanatory use of the expression model: an incorrect alignment would (at the misaligned passages) lead to an explanation of the loudness in terms of passages of the score that are not actually being played.

\subsection{From solo to ensemble performance}\label{sec:from-solo-ensemble}
In terms of modeling, there are several significant differences between a solo instrument setting and an ensemble setting. Firstly, in an ensemble setting, multiple sets of basis-functions are produced, each set describing the score part of a particular instrument.
Furthermore, in a symphonic piece, multiple instantiations of the same instrument may be present.
Lastly, different pieces may have different instrumentations.
This poses a challenge to an expression model, which should account for the influence of instruments consistently from one piece to the other.
We address these issues by defining a \emph{merging} operation that combines the information of different sets of basis-functions for each instance of an instrument into a single set of basis-functions per instrument class.

The way dynamics is measured and represented (see Section~\ref{sec:meas-dynam-music}) also has repercussions for the basis-function modeling approach.
In contrast to the digital grand piano setting, the overall loudness measured from an orchestral recording does not provide a loudness value for each performed note, but one per time instant.
Thus, basis-function information describing multiple simultaneous notes must be combined to account for a single loudness value.
We do so by defining \emph{fusion} operators for subsets of basis-functions.
In most cases, we use the \emph{average} operator as a default.
For some basis-functions however, we use the \emph{sum} operator, in order to preserve information about the number of instances that were fused into a single instrument.
Future experimentation should provide more informed choices as to the optimal fusion operators to use.

\begin{figure*}[t]
\centering
\ignorewhencounting{
  \includegraphics[width=0.95\linewidth]{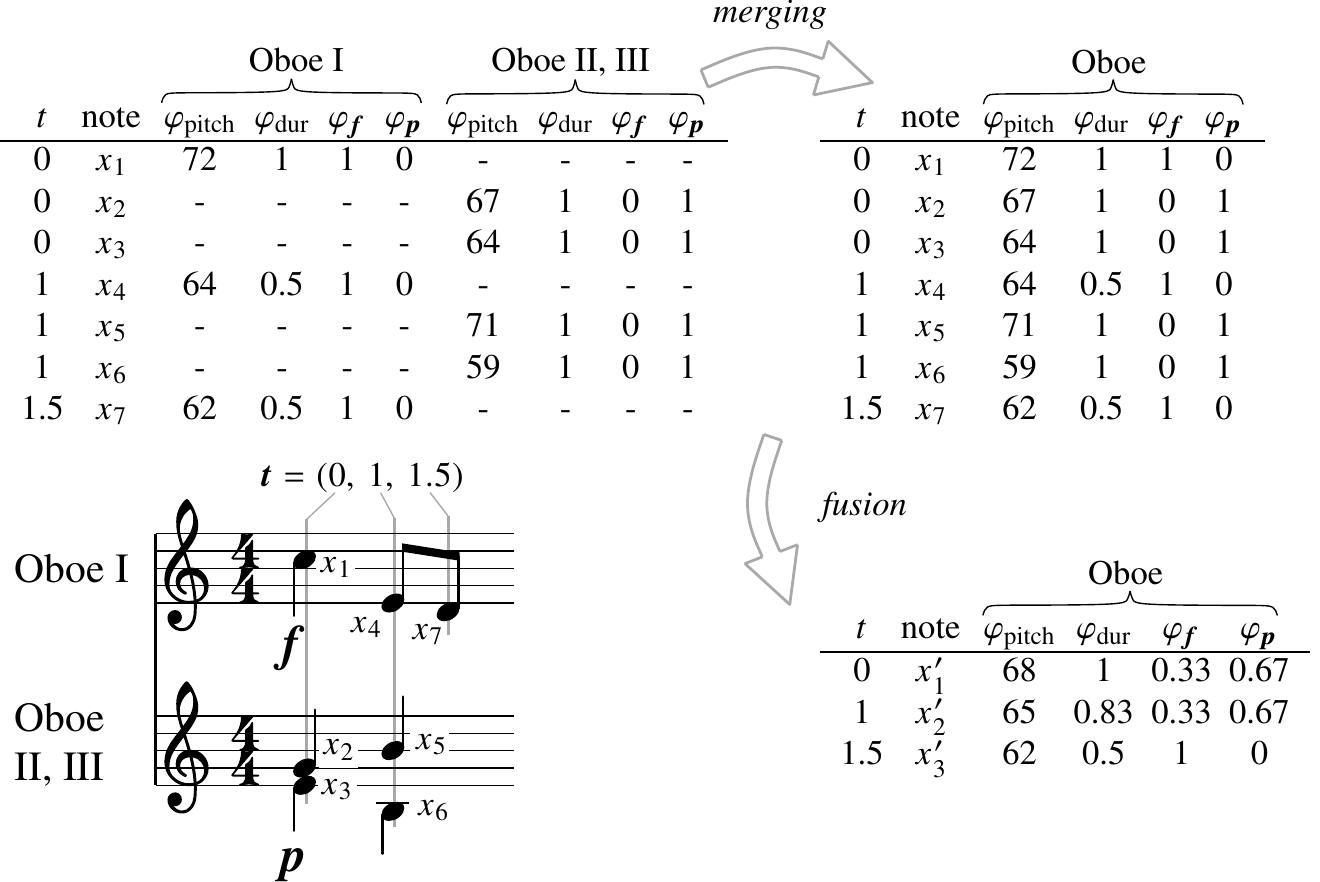}

\caption{Illustration of \emph{merging} and \emph{fusion} of score information of two different parts belonging to the same instrument class \enquote{Oboe}.
The first matrix shows four example basis-functions, $\varphi_{\mathrm{pitch}}$, $\varphi_{\mathrm{dur}}$, $\varphi_{\vex{f}}$ and $\varphi_{\vex{p}}$, for the notes of each of the two score parts.
The second matrix is the result of \emph{merging} basis-functions of different \enquote{Oboe} instantiations into a single set.
The last matrix is the result of \emph{fusion}, applied per basis-function to each set of values occurring at the same time point}
}
\label{fig:merge_and_fuse}
\end{figure*}

Both the merging and the fusion operations are illustrated for a musical excerpt in Figure~\ref{fig:merge_and_fuse}.

\subsection{Linear, non-linear, and temporal models}
\sloppypar
Initial versions of the basis-function expression model used a linear model (Lin)~\citep{grachten12:_linear}.
In a linear model, the expressive parameters are simply a weighted sum of the basis-functions, where the parameters of the model are the weights for each basis-function, to be estimated based on training data.
A major advantage of a linear model is that the link between the basis-functions and the predictions is very clear: the weight for a basis-function expresses how strongly the basis-function influences the output.
This makes it easy to perform a qualitative analysis of what the model has learned, and by fitting the model on a particular piece, or on several pieces by the same performer, the weights may also capture characteristics of the expressive quality of a piece, or a performer \citep{grachten12:_linear}.

The simplicity of linear modeling is at the same time a drawback.
There are two main limitations to the linear approach.
Firstly, the shape of a basis-function can only be used literally (apart from scaling and vertical translation) to approximate an expressive parameter.
For example, a \emph{crescendo} annotation is schematically represented as a \emph{ramp function}, and this means that any increase of loudness in that region can only be approximated as a linear slope.
In reality, it is likely that the shape of the loudness increase is not strictly linear.
Secondly, the linear approach does not model any interactions between basis-functions.

\sloppypar
To overcome these limitations, \citet{chacon15} proposed a non-linear basis-function model for expression, based on a \emph{feed-forward neural network} (FFNN), showing the advantages over a linearity of the model, both in the non-linear transformation of the basis-functions, and in the interaction between basis-functions.

A more powerful type of non-linear modeling can be obtained by introducing recurrence relations to the neural network architecture:
\emph{recurrent neural networks} (RNNs) are a particular kind of discrete--time dynamical artificial neural networks suited for analyzing sequential data, such as time-series.
These dynamic models have been successfully used for generating text sequences, handwriting synthesis and modeling motion capture data~\citep{Graves:2013sm}.
The structure of an RNN is similar to that of a feed forward neural network, with the addition of connections among its subsequent hidden states, allowing information from the past to influence the hidden state that corresponds to the present.

\begin{figure}[t]
  \begin{center}
\ignorewhencounting{
  \includegraphics[width=.7\textwidth]{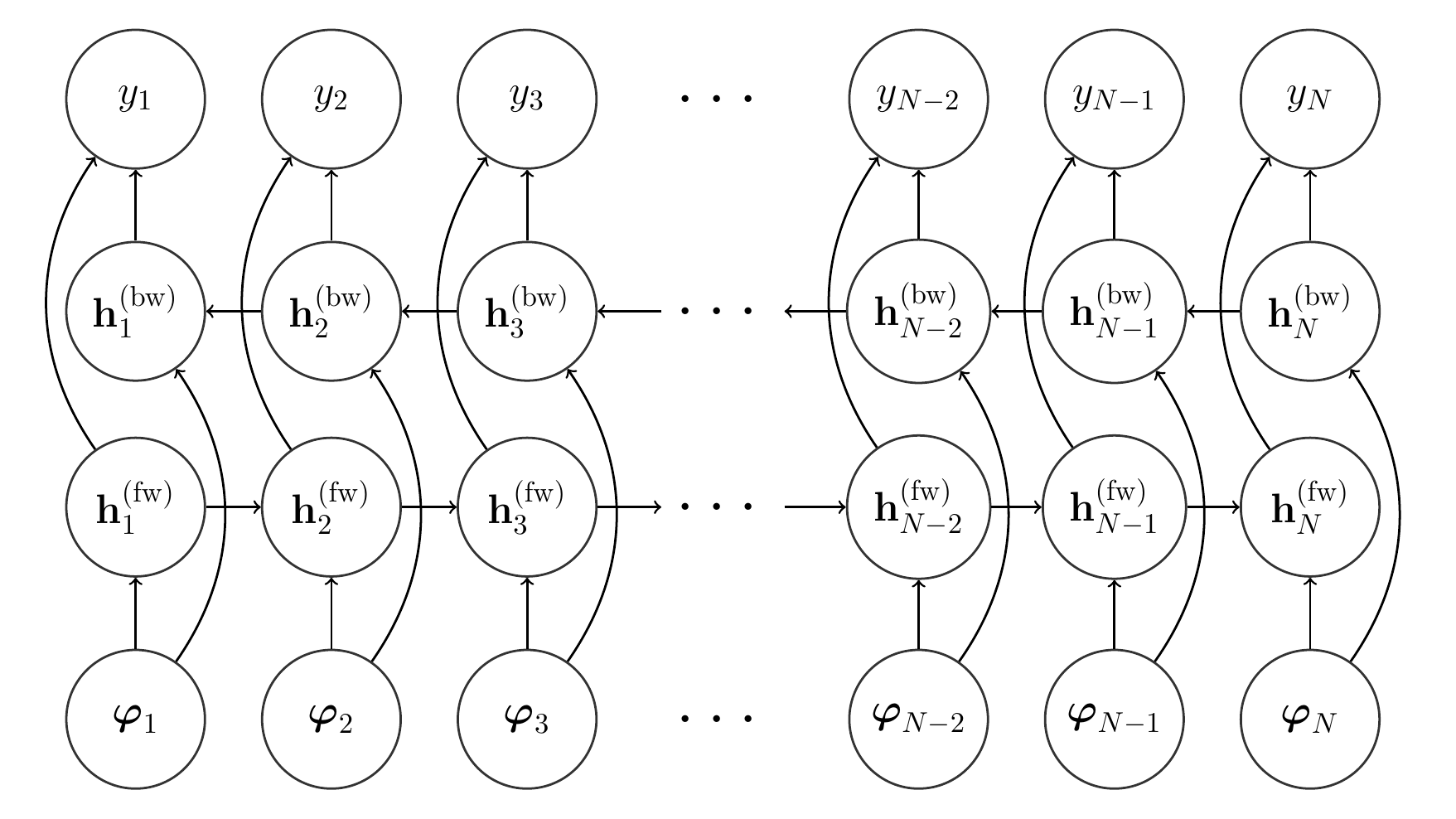}
    \caption{Graph structure of a biRNN, expanded for a piece of $N$ time steps. $\vex{\varphi}_t$ is a vector holding the values of a set of basis-functions $\vex{\varphi}$, evaluated at time $t$, describing the musical score at that time. $\mathbf{h}^{(\mathrm{fw})}_t$ and $\mathbf{h}^{(\mathrm{bw})}_t$ model the forward and backward score context at $t$, respectively, and jointly predict $y_t$, the performance dynamics at $t$}
  }
    \label{fig:birnn-arch}
  \end{center}
\end{figure}

RNNs are not limited to forward temporal dependencies: Correlations between present and future events may be modeled by \emph{backward} temporal connections.
An RNN with both forward and backward connections is referred to as a \emph{bi-directional} RNN (biRNN).
The benefits of such models have been demonstrated in the context of expressive timing variations in Chopin piano music~\citep{grachten16:brnbm}.
Figure~\ref{fig:birnn-arch} illustrates how the biRNN is used to predict the dynamics of a performed piece, given the basis-function description of the piece $\vex{\varphi}$.
The graph structure of a FFNN is similar, but lacks the lateral connections between hidden states $\mathbf{h}$, and only has a single hidden layer $\mathbf{h}$, replacing $\mathbf{h}^{(\mathrm{fw})}_t$ and $\mathbf{h}^{(\mathrm{bw})}_t$.

\section{An experimental assessment of the model}
In this Section, we provide an evaluation of the model in terms of \emph{predictive accuracy} on a set of orchestra performances that are available commercially.
Such an evaluation may not at first seem to be of major interest, since we intend to use the expression model as a tool for \emph{understanding} expression rather than predicting it.
However, it should be kept in mind that in the area of musical expression, there is still little consolidated theory to use as a basis for building models.
Therefore, computational models of musical expression by necessity have an exploratory role.
In this context, measuring the predictive accuracy of the model on a set of music recordings helps us to get a general impression of how well the model captures relevant factors of the music in relation to expression.
By testing both simple linear and more complex associations between the basis-function representation and the recorded loudness of performances, we aim to give a more complete picture of the merits and limitations of the basis-function modeling approach.
More specifically, we test the linear basis-function model (Lin), the non-linear model (FFNN), and the bi-directional recurrent model (biRNN) described in the previous Section.

\subsection{Data}\label{sec:data}
The corpus used for the experiments is summarized in Table \ref{table:data} below.
It consists of symphonies from the classic and romantic period.
The corpus contains recorded performances (audio), machine-readable representations of the musical score (MusicXML) and automatically produced, manually corrected alignments between score and performance, for each of the symphonies.

\begin{table}[h]
\footnotesize
\centering
\ignorewhencounting{
\caption{Pieces/movements used for a leave-one-out cross-validation of the model}
\begin{tabular}{ccccc}
\toprule
Composer & Piece & Movements & Conductor & Orchestra \\
\midrule
Beethoven & Symphony No.
5 in C-Min.
(op. 
67)
& 1, 2, 3, 4 &Fischer & RCO \\
Beethoven & Symphony No.
6 in F-Maj.
(op.
68)  & 1, 2, 3, 4, 5 &Fischer & RCO \\
Beethoven & Symphony No.
9 in D-Min.
(op.
125) & 1, 2, 3, 4    &Fischer & RCO \\
Mahler    & Symphony No.
4 in G-Maj.
          & 1, 2, 3, 4    &Jansons & RCO \\
Bruckner  & Symphony No.
9 in D-Min.
(WAB 109) & 1, 2, 3       &Jansons & RCO \\
\bottomrule
\end{tabular}
}
\label{table:data}
\end{table}

We use recordings of performances by the Royal Concertgebouw Orchestra conducted by Ivan Fischer or Mariss Jansons, all performed at the Royal Concertgebouw in Amsterdam, the Netherlands.
The corpus amounts to a total of 16 movements from 4 pieces.
The corresponding performances sum up to a total length of almost 4 hours of music.
From the 20 scores a total of 53816 note onsets, and 1420 basis-functions were extracted.
The loudness and score--performance alignment is computed as described in Section~\ref{sec:meas-dynam-music}.

\subsection{Method}\label{sec:method}
We use a leave-one-out cross-validation, where the model is trained on 19 of the 20 movements and then is used to predict the target values for the unseen remaining movement.
The non-linear models (FFNN, and biRNN) are trained by gradient descent optimization.
Both the feed-forward and the recurrent neural network are set up with a single hidden layer of 20 units.
From the 19 training movements, four movements are kept for validation, to avoid overfitting the models to the training data, a practice known as \emph{early stopping}. 
The predictions are evaluated with respect to the target in terms of the Coefficient of Determination ($R^2$), measuring the proportion of variance explained by the model, and Pearson's Correlation Coefficient $r$. Note that since we report on predicted rather than fitted data, $R^2$ values can be negative, in case the prediction residual has a larger variance than the signal itself.

The set of basis-functions used in the experiments encode note pitch, duration, and metrical position, the number of simultaneous notes within instrument groups, inter-onset intervals between subsequent notes, repeat signs, note accent, staccato, fermata signs, and dynamic markings.
A full description of the basis-functions is omitted for brevity\footnote{A complete description of the basis functions can be found in the following technical report: \url{http://lrn2cre8.ofai.at/expression-models/TR2016-ensemble-expression.pdf}}.

\subsection{Results and discussion}\label{sec:results-discussion}
The results of the experiments are shown in Table~\ref{table:results}.
We observe that both the $R^2$ and the $r$ values for Lin are generally lower than those for FFNN and biRNN, demonstrating that the non-linear modeling provides a clear advantage over the linear modeling approach.
Given the relatively small data set, this result is not trivial, since the FFNN and biRNN have much more parameters than the Lin model, and are therefore more prone to \emph{overfitting}.
Furthermore, the biRNN model provides more accurate predictions than the FFNN model, although this advantage is less prominent than the advantage over Lin.


\begin{table*}[]
\footnotesize
\ignorewhencounting{
  \caption{Predictive accuracy in a leave-one-out scenario for the different models;
$R^{2}$ = coefficient of determination (larger is better); $r$ = Pearson correlation coefficient (larger is better);
Best value per piece and measure emphasized in bold}
\label{table:results}
\newcommand{\inw}{\hskip1em}
\newcommand{\outw}{\hskip2.8em}
\newcommand{\thead}[2]{\multicolumn{1}{c@{#1}}{#2}}
\begin{center}
\begin{tabular}{
l@{\hskip.4em}l@{\outw}
r@{\inw}r@{\inw}r@{\outw}
r@{\inw}r@{\inw}r@{\outw}
r@{\inw}r@{\inw}r}
\toprule
Composer / Piece & &
\multicolumn{3}{c@{\outw}}{$R^2$} &
\multicolumn{3}{c@{\outw}}{$r$} \\

& &
\thead{\inw}{Lin} & \thead{\inw}{FFNN} & \thead{\outw}{biRNN} &
\thead{\inw}{Lin} & \thead{\inw}{FFNN} & \thead{\outw}{biRNN} \\
\midrule



Beethoven S5 & Mv 1 & -0.26  & -0.22 & \textbf{-0.18} & 0.18 & 0.21 & \textbf{0.26} \\
             & Mv 2 & 0.34   & 0.46  & \textbf{0.56}  & 0.58 & 0.70  & \textbf{0.76} \\
             & Mv 3 & 0.23   & 0.40   & \textbf{0.44}  & 0.53 & 0.64 & \textbf{0.66} \\
             & Mv 4 & 0.06   & \textbf{0.26}  & 0.25  & 0.41 & \textbf{0.53} & 0.52 \\
Beethoven S6 & Mv 1 & 0.36   & 0.36  & \textbf{0.39}  & 0.61 & 0.63 & \textbf{0.65} \\
             & Mv 2 & 0.07   & 0.15  & \textbf{0.17}  & 0.36 & 0.40  & \textbf{0.41} \\
             & Mv 3 & 0.51   & 0.60   & \textbf{0.62}  & 0.72 & 0.81 & \textbf{0.82} \\
             & Mv 4 & 0.11   & 0.27  & \textbf{0.29}  & 0.38 & 0.54 & \textbf{0.56} \\
             & Mv 5 & 0.36   & 0.44  & \textbf{0.49}  & 0.60  & 0.70  & \textbf{0.75} \\
Beethoven S9 & Mv 1 & 0.34   & 0.36  & \textbf{0.42}  & 0.59 & 0.61 & \textbf{0.65} \\
             & Mv 2 & 0.36   & 0.40   & \textbf{0.53}  & 0.60  & 0.64 & \textbf{0.74} \\
             & Mv 3 & -0.30   & -0.06 & \textbf{-0.02} & 0.20  & 0.17 & \textbf{0.22} \\
             & Mv 4 & 0.11   & 0.37  & \textbf{0.49}  & 0.52 & 0.64 & \textbf{0.70}  \\
Mahler S4    & Mv 1 & -0.17  & 0.29  & \textbf{0.37}  & 0.37 & 0.54 & \textbf{0.61} \\
             & Mv 2 & -0.48  & -0.02 & \textbf{-0.02} & 0.06 & 0.20  & \textbf{0.23} \\
             & Mv 3 & -1.22  & 0.25  & \textbf{0.26}  & 0.20  & 0.51 & \textbf{0.53} \\
             & Mv 4 & -1.99  & 0.09  & \textbf{0.18}  & 0.15 & 0.33 & \textbf{0.44} \\
Bruckner S9  & Mv 1 & -39.06 & 0.45  & \textbf{0.59}  & 0.26 & 0.68 & \textbf{0.77} \\
             & Mv 2 & 0.24   & 0.48  & \textbf{0.55}  & 0.58 & 0.72 & \textbf{0.74} \\
             & Mv 3 & -3.54  & 0.32  & \textbf{0.40}   & 0.25 & 0.57 & \textbf{0.65} \\

\bottomrule
\end{tabular}
\end{center}
}
\end{table*}

The fact that the results of the linear model are inferior suggests that although the basis-functions used to represent the score capture relevant information, their shapes (such as the ramp function to represent a \emph{crescendo}) are too schematic to work well as approximations of measured loudness curves.
The improvement of the results in the FFNN and RNN models suggest that the non-linear transformation of these shapes alleviates this problem to some extent.
The capability of the non-linear models of modeling interactions between basis-functions, as demonstrated by \cite{chacon15}, may further explain the improved results of these models.

\section{The expression model as an analytical tool for dynamics in symphonic music}\label{sec:expression-model-as}

In this Section, we demonstrate that the BM framework can be used for explanatory purposes, and thus form the basis for a tool that elucidates differences in expressive interpretations between performances, as discussed in Section~\ref{sec:introduction}.

The explanatory power of BM models lies in the fact that they represent dynamics as a
function of the basis-functions.
As a model learns from training data how the basis-functions relate to dynamics, some basis-functions may prove to be very important for an accurate prediction of dynamics, while others may have little or no influence at all.
In other words, the model learns to be more \emph{sensitive} to some 
basis-functions than to others.
We can impose sensitivities specific to a particular performance on a model by \emph{fitting} the model to that performance---adjusting its parameters such that its prediction error for the dynamics of that performance is minimized.
When fitting models to two different performances of a piece, the differences in dynamics between the performances tend to lead to different sensitivities in the models.
For example, a model fitted to a dramatic performance may learn that dynamics annotations such as \emph{piano} and \emph{forte} have a large effect on the dynamics of the performance, whereas a model fitted to a more restrained performance may be less sensitive to these annotations.
Thus, comparing differences in sensitivities between models fitted to different performances can give us qualitative explanations of the differences in dynamics, in the style of ``Performance A is louder than performance B at this point in the piece, because the string instruments are more prominent'', or ``Performance B emphasizes the downbeat more strongly than Performance A''.

When fitting two models to two different performances for comparison purposes, it makes sense to start the fitting process from a common model that was \emph{pretrained} on a number of other recordings.
Firstly, this speeds up the fitting process since the pretrained model will already provide a rough approximation of the dynamics curves, and secondly, starting from a common basis encourages similar explanations for similar trends in the performances, and thereby parsimonious explanations of the differences between the performances.

We compute the sensitivities of a model to each of the basis-functions using a \emph{local differential--based sensitivity analysis} technique~\citep{hamby1995}, which consists in computing the gradient of the output of the model with respect to each of its inputs.
By multiplying the gradients (sensitivities) with the inputs (basis-functions) over the course of a piece, we obtain a \emph{sensitivity graph} for a performance.
The multiplication is motivated by the fact that even when a model is sensitive to a particular basis-function, this basis-function does not affect the output of the model whenever it is inactive (i.e. zero valued).
Moreover, by subtracting the sensitivity graphs of two performances, we obtain what we call a \emph{sensitivity-difference (SD) graph}, a visual representation expressing the relative influence of each basis-function in each of the two performances, to be illustrated shortly.




As a case study, we compare performances of the 3rd Movement (\textit{Lustiges Zusammensein der Landleute}) of Beethoven's Symphony No.
6 Op.
68 by different conductors and orchestras.
This piece is a scherzo which suggests dances of the country folk.
The scherzo is in a \setmeter{3}{4} meter, with its trio in a \setmeter{2}{4} meter.
The two performances we compare here are by the conductors Georg Solti (with the Chicago Symphony Orchestra, recorded in 1974) and Nikolaus Harnoncourt (with the Chamber Orchestra of Europe, recorded in 1991), hereafter referred to as Solti and Harnoncourt, respectively.
We compare the performances by using a biRNN model that was pretrained on the dataset described in Table~\ref{table:data}, and then fitting the pretrained model to both performances, respectively.


\begin{figure}[t]
  \begin{center}
    \ignorewhencounting{
\includegraphics[width=\linewidth]{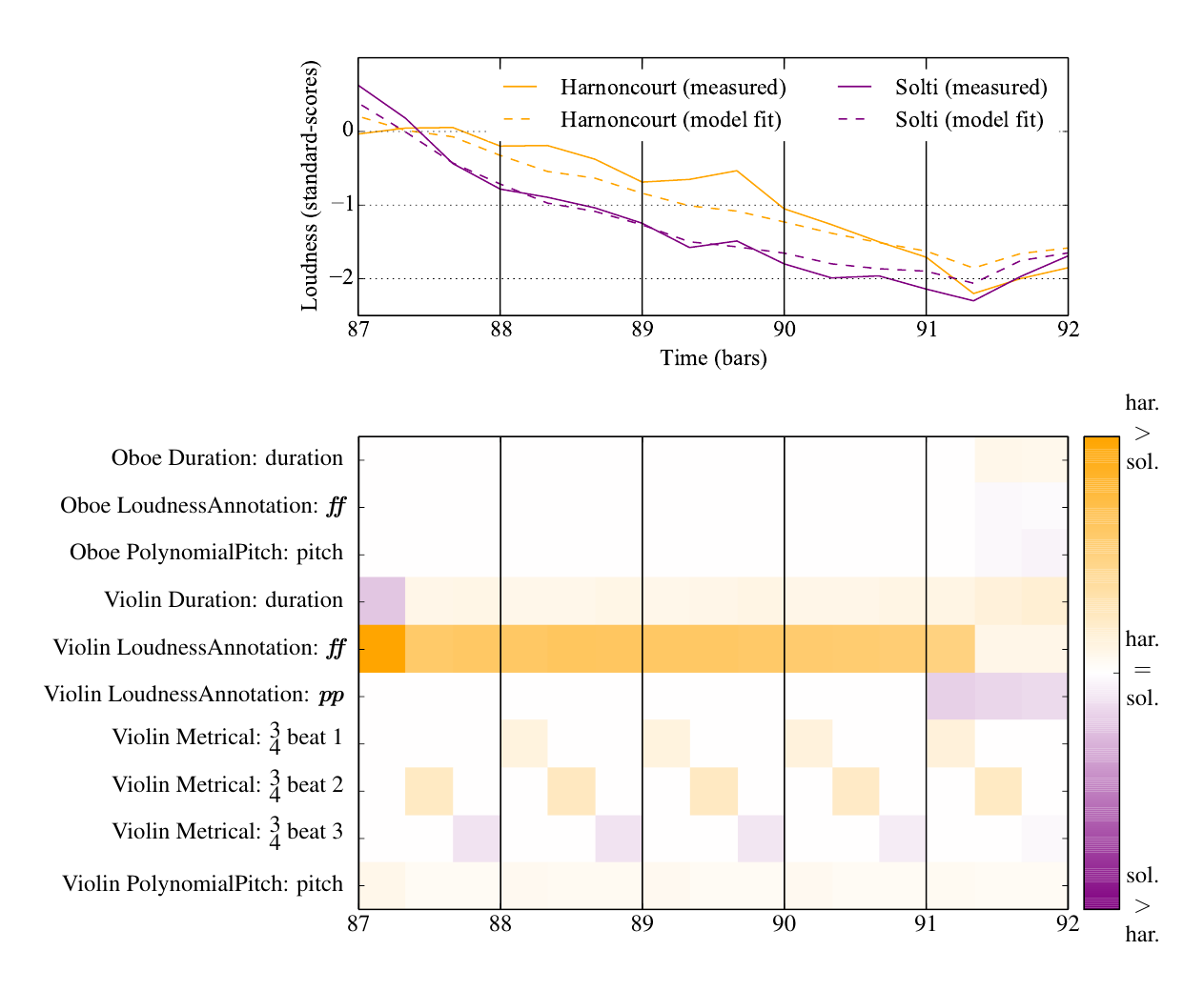}
\caption{Top: Measured and fitted loudness curves for an excerpt of the performance of Beethoven's 6th Symphony, 3rd Movement (bars 87 to 92) by Harnoncourt (orange) and Solti (purple); Bottom: sensitivity-difference graph for Harnoncourt and Solti. Orange tones indicate that a basis-function has a stronger (positive) contribution to loudness in Harnoncourt than in Solti, purple tones indicate the opposite. (Figure best viewed in color)}
}

\label{fig:solti-vs-harnoncourt}
\end{center}
\end{figure}

\sloppypar
As an example, consider the small, but marked difference between Solti and Harnoncourt in bars 87-90 of the SD graph in Figure~\ref{fig:solti-vs-harnoncourt} (showing a selection of the most influential basis-functions in the fragment).
In bar 87, Beethoven's score prescribes a four-bar \emph{diminuendo} of the violins to transition from an ongoing \emph{fortissimo} (\emph{ff}) passage (starting before and continuing in the depicted fragment) to a quiet and lyrical \emph{pianissimo} (\emph{pp}) passage featuring a singing oboe, starting with bar 91.
The SD graph shows that the increased loudness in Harnoncourt (compared to Solti) is attributed to a sustained influence of the \emph{ff} in the violins over the course of the diminuendo.
Note that this attribution is a parsimonious explanation of the loudness difference, because it involves only a single basis-function.
A hypothetical, less concise explanation for instance, could involve an increased influence of each of the metrical positions for Harnoncourt.
Together, the increased influence of these basis-functions would also lead to a louder performance of the fragment overall, but may not be compatible with Harnoncourt's interpretation of the rest of the piece.

Listening to the respective passages, we note indeed a clearly audible difference:
Solti takes the \emph{diminuendo} very strictly, immediately softening the orchestra and quickly arriving at a very soft playing level already before the actual arrival of the \emph{pp}.
Harnoncourt's \emph{ritardando} is more of a continuation of the preceding \emph{fortissimo} passage: he only grows slightly softer during the \emph{ritardando}, and obeys the \emph{pp} more abruptly when it arrives (the purple color of the \emph{pp} starting with bar 91 indicates that Solti's \emph{pp} is actually slightly louder than Harnoncourt's).
It turns out that these are consistent and obviously deliberate choices, as we find the exact same pattern later on in the piece, in bars 292--295, where we have an analogous musical passage.

Furthermore, the SD graph shows a slight but systematic pattern in the metrical basis-functions.
This pattern suggests the model found slight differences in the metrical accentuation, with Solti placing more emphasis than Harnoncourt on the last beat of the bar, and vice versa for the first to beats.
Listening reveals that these differences are too subtle to be heard, however.

Finally, it is important to note that the SD graph pertains to the \emph{model fit} dynamics curves in Figure~\ref{fig:solti-vs-harnoncourt} (top), not the \emph{measured} curves. There are some fluctuations in the measured curve (such as that on beat 3 of measure 89) that are not captured, and therefore cannot be explained by the SD graph.

\section{Conclusions}
This article demonstrates a computational model for modeling loudness variations in audio recordings of symphonic pieces.
The model relates these variations to information from the written musical score, described in the form of \emph{basis-functions}.
An evaluation of different variants of the model shows that a non-linear version including temporal dependencies is most effective in a predictive setting, where the model predicts loudness variations based on the written score, after being trained on a set of recordings.
Examples given in Section~\ref{sec:expression-model-as} illustrate how the model can be used as a way to explain differences between performances in terms of the written score.

It must be kept in mind however, that the data set used for validating the model, although comprising works from several composers, is performed by a single orchestra, and two different conductors. Anecdotal cross-validation suggests the model trained on RCO recordings generalizes well to recordings by different orchestras, but more elaborate experimentation is necessary to make stronger claims about the robustness of the model against variance in recording/mixing/mastering conditions across recordings.

Furthermore, the measured overall loudness variation is only a coarse measure of (a single aspect of) musical expression, and that currently, the model approximations (and thus its explanations) may not be adequate at all positions in the performance.
Better model approximations and predictions will allow for novel explanatory uses, such as using the predictions of a model that was trained on multiple performances of a piece as a ``baseline'' performance, based on which the idiosyncrasies of conductors can be established.

The examples were hand-picked here, since the expression model is currently in a stage where we are testing its validity, and experimenting with different sets of basis-functions.
In the future, the model should be capable of automatically identifying excerpts from a piece where two or more performance differ substantially from each other, in order to highlight them to the listener, and show which aspects of the performance are different.
Future versions of the model should not be restricted to loudness variations, but cover tempo variations as well.

In combination with a web-service for aligned music playback and visualization, such as presented by~\cite{gasser15:_class_music_web_user_inter_data_repres}, the model presented here allows listeners with a desire to get a better grasp on a piece of music, to compare different performances of the piece in terms of their expressive character, and get a better understanding of what it is that makes the performances different.

\section*{Acknowledgements}
This work is supported by the European Union's Seventh Framework Programme FP7 / 2007-2013 (projects PHENICX / grant number 601166 and Lrn2Cre8 / grant number 610859), and by the European Research Council (ERC) under the EU's Horizon 2020 Framework Programme (ERC Grant Agreement number 670035, project CON ESPRESSIONE).
We wish to thank the anonymous reviewers for their useful comments.
Furthermore, we are grateful to the Royal Concertgebouw Orchestra, in particular Marcel van Tilburg and David Bazen, for providing the audio recordings used in this study.

\bibliographystyle{apalike}

\IfFileExists{bibliographies/bib_mg.bib}{
\bibliography{bibliographies/bib_mg,bibliographies/bib_cc,bib_tg}
}{
\bibliography{bib_mg,bib_tg}
}

\section*{Author Bibliographies}
\paragraph{Maarten Grachten}
holds a Ph.D. degree in Computer Science and Digital Communication (2007, Pompeu Fabra University, Spain), and is currently senior researcher at the Department of Computational Perception at Johannes Kepler University, Linz, Austria. His research interests include computational modeling of music perception and cognition using machine learning techniques.

\paragraph{Carlos Eduardo Cancino-Chac\'on}
is a researcher at the Austrian Research Institute for Artificial Intelligence (OFAI). He received the Bachelor's degree in Physics at UNAM, Mexico, the Bachelor's degree in Piano at the National Conservatory of Music in Mexico and the MSc. in Electrical Engineering and Audio Engineering at TU Graz, Austria. Currently he is pursuing the PhD degree in Computer Science at JKU, Linz, Austria.

\paragraph{Thassilo Gadermaier}
received a Mag. phil. degree in Systematic Musicology from University of Vienna and is currently a researcher at OFAI. He is pursuing a degree in Electrical Engineering/Telecommunications at Vienna University of Technology.

\paragraph{Gerhard Widmer}
is Professor and Head of the Department of Computational
Perception at Johannes Kepler University, Linz, Austria.
His research interests include AI, machine learning, and intelligent
music processing. He is a Fellow of the European Association for Artificial
Intelligence, and recipient of an ERC Advanced Grant (2015)
on computational models of expressivity in music.

\end{document}